\documentclass{rspublic}
\usepackage{amsmath,amsthm,amssymb}
\usepackage[numbers, sort&compress]{natbib}
\usepackage{graphicx}

\newcommand{\figref}[1]{Figure~\ref{#1}}
\newcommand{\figrefp}[1]{(\figref{#1})}
\newcommand{\ddt}[1]{\frac{\mathrm{d}#1}{\mathrm{d}t}}

\author{Michel Crucifix}
\affiliation{Georges Lema\^\i tre Centre for Earth and Climate Research \\ 
Earth and Life Institute \\
Universit\'e catholique de Louvain}
\title{Oscillators and relaxation phenomena in Pleistocene climate theory}
\begin{document}
\bibliographystyle{rsociety}
\maketitle
\bigskip
\begin{abstract}{palaeoclimates, dynamical systems, limit cycle, ice ages, Dansgaard-Oeschger events}
Ice sheets appeared in the northern hemisphere around 3 million years ago and glacial-interglacial cycles have paced Earth's climate since then.  Superimposed on these long glacial cycles comes an intricate pattern of millennial and sub-millennial variability, including Dansgaard-Oeschger and Heinrich events. 

There are numerous theories about theses oscillations. Here, we review a number of them in order to draw a parallel between climatic concepts and dynamical system concepts, including, in particular, the relaxation oscillator, excitability, slow-fast dynamics and homoclinic orbits. 

Namely, almost all theories  of ice ages reviewed here feature a phenomenon of synchronisation between internal climate dynamics and the astronomical forcing. However, these theories differ in their bifurcation structure and this has an effect on the way the ice age phenomenon could grow 3 million years ago.
All theories on rapid events reviewed here rely on the concept of a limit cycle, which may be excited by changes in the ocean surface freshwater balance. 
The article also reviews basic effects of stochastic fluctuations on these models, including the phenomenon of phase dispersion, shortening of the limit cycle and stochastic resonance. 
It concludes with a more personal statement about the potential for inference with simple stochastic dynamical systems in palaeoclimate science. 
\end{abstract}
\section{Introduction}
\begin{figure}[t]
\begin{center}
\includegraphics[width=\textwidth]{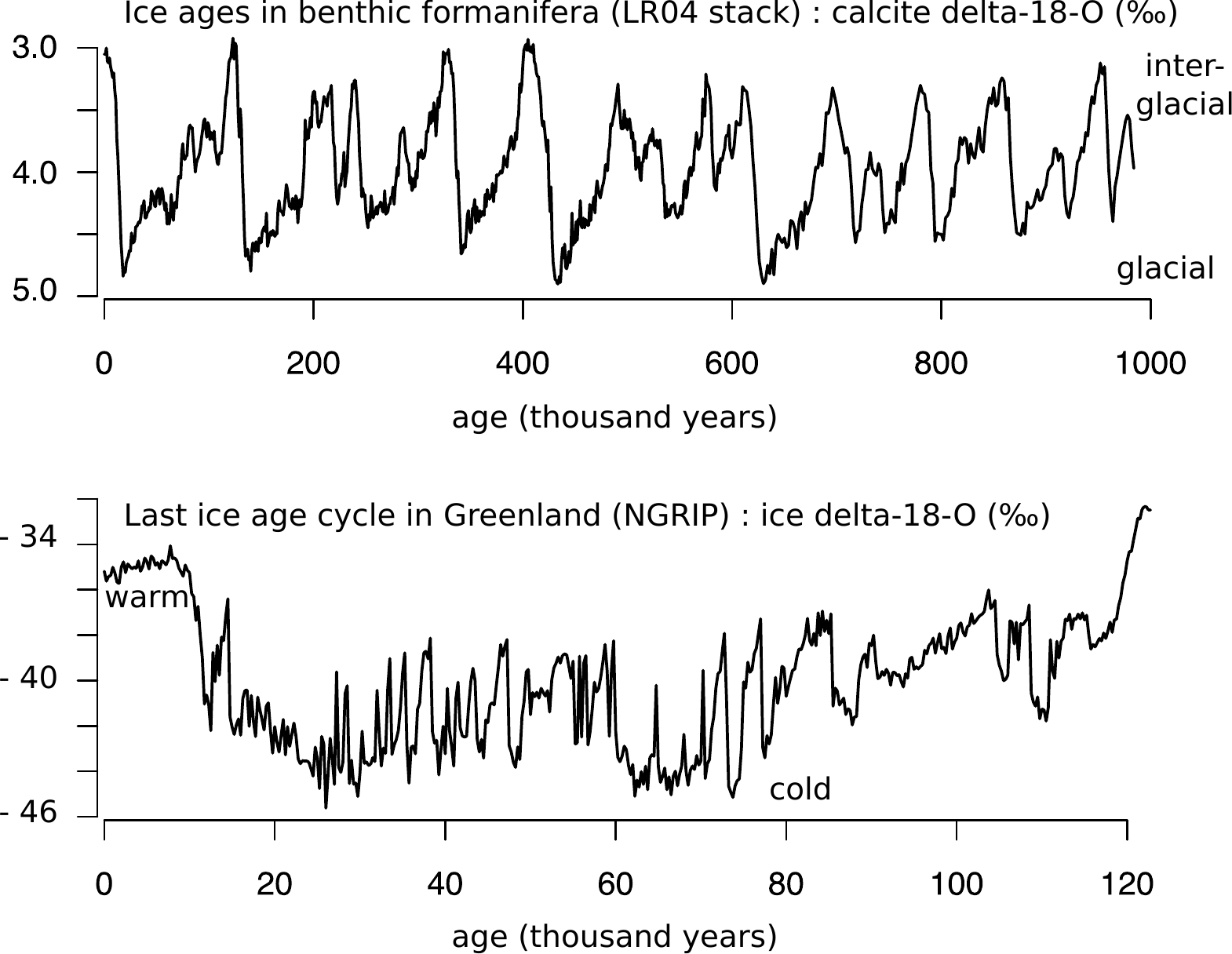}
\end{center}
\caption{Climatic fluctuations over the late Pleistocene. Ice ages are reconstructed using the oxygen isotopic ratio of the calcite shells of benthic foraminifera \citep{lisiecki05lr04}. Within the last ice ages, large temperature fluctuations were recorded in Greenland, here depicted by changes in the oxygen isotopic ratio of ice \citep{GRIP04aa}}
\label{fig:data}
\end{figure}

The Pliocene and the Pleistocene cover approximately the past five million years. 
The climatic fluctuations that characterized this period may be reconstructed from numerous natural archives, including marine, continental and ice core records. These archives show a complex climate history. Ice sheets appeared in the northern hemisphere around 3 to 3.5 million years ago \citep{Shackleton84ab,Meyers10aa}. The volume of these ice sheets fluctuated with the variations of the seasonal and spatial distributions of incoming solar radiation (insolation), which are induced by changes in the geometry of the Earth's orbit and the angle (obliquity) between  Earth's equator and the ecliptic \citep{croll1875,  milankovitch41, berger78}. This is called the astronomical forcing
\footnote{The astronomical forcing will generally be taken into account here in the form of a normalised measure of insolation during the month or on the day of summer solstice at a northerly latitude, typically 60 or 65$^\circ$ N. This is a fairly complex, aperiodic signal, with dominant harmonics corresponding to the phenomena of precession (23716,  22428 and 18976 years); and obliquity (41000 years) \citep{berger78}}. 
Glacial cycles had an average duration of about 40,000 years \cite{ruddiman86} until about 800,000 years ago. 
The dominant period of glacial cycles increased around 800,000 years ago and this is referred to as the Middle Pleistocene Transition. Data and models about the Middle Pleistocene Transition are reviewed in ref. \citep{Clark06aa}. Time-series analyses based on band-pass filtering provide further evidence of the non-linear nature of the climate response to the astronomical forcing, from  about  1.4 Myr ago \citep{lisiecki07trends}. 
The latest four glacial cycles, in particular, are distinguished by a pronounced saw-tooth time-structure: ice accumulates over the continents during about 80,000 years and then melts in about 10,000 years \figrefp{fig:data}. 

Superimposed on these long glacial cycles comes a complex pattern of millennial and sub-millennial variability \citep{McManus99aa}. For example, the Greenland record features at least 20 events of abrupt rise and slower decline in oxygen isotopic ratio (a proxy for temperature) \citep{johnsen92aa,dansgaard93} and methane \citep{Chappellaz93aa} during the latest glacial epoch. These events are known as Dansgaard-Oeschger events. They were found to occur from  at least the last glacial inception  \citep{Capron10aa} and the Antarctic ice core record provide evidence that they are characteristic of Pleistocene glacial climates \citep{Loulergue08aa}.  
Some of these events follow pulses of iceberg discharges into the North Atlantic Ocean, called `Heinrich events' \citep{heinrich88, bond92heinrich,  grousset93ird}. Heinrich events and Dansgaard-Oeschger events have left climatic footprints all over the globe \citep{Voelker02aa}, including in Antarctica \citep{Loulergue08aa}.
The current interglacial period is referred to as the Holocene. It is also characterised by millennial and centennial variability, mainly observed in the North Atantic \citep{Bond97aa,Bianchi99aa,Paul02aa}, but of a much weaker amplitude than during the preceding glacial period. 

The present paper reviews attempts to explain these fluctuations with concepts that originate in dynamical system theory. These are the concepts of limit cycle, synchronisation and excitability. The central message of the paper is that current theories of ice ages and rapid events may often be interpreted in terms of generic deterministic models, which are also used in other areas of Science like biology and ecology. However, stochastic parameterisations are an essential part of any complex system model, and their effects on climatic oscillations have to be taken into account. 

Dynamical system theory entered palaeoclimate science with idealised models representing the response of ice sheets to the astronomical forcing. These models were directly derived from the physics of the ice-sheet-atmosphere system \citep{oerlemans80,  OERLEMANS82aa, ghil81, letreut83}. Ghil and Childress \cite{Ghil87aa}, in particular, insisted on the interest of analysing such models in terms of bifurcation theory. For modelling the complex carbon cycle response authors sometimes adopted a more heuristic approach by considering simple models and confronting the results to palaeoclimate evidence \citep{saltzman88}.

Nowadays climate research is largely oriented towards large climate simulators (typically: general circulation models), which are developed to include as many climate processes as possible. However, thinking in terms of dynamical system theory remains insightful. Indeed, the behaviour of a complex system at a certain spatio-temporal scale is in practice often dominated by a few leading modes, of which the dynamics may be captured fairly convincingly with a low-order dynamical system. Climate scientists are increasingly using this property. For example, they formulate simpler models to explain the seemingly complex behaviours observed in ocean-atmosphere simulators. 
Examples have been provided in the recent years focusing on interannual \citep{Timmermann03aa}, centennial \citep{Schulz07aa} and millennial \citep{Sakai99aa, Colin-de-Verdiere07aa} variability. In parallel, so-called hysteresis experiments, which aim at identifying the number of stable states in individual components of the climate system such as the ocean circulation \citep{Rahmstorf05aa} or ice sheets \citep{Calov05aa} contribute to a dynamical-system founded understanding  of the climate system. 
This approach may also help us to predict and communicate about the proximity of bifurcations, which may result in catastrophic climatic changes.  Timmermann and Jin \citep{Timmermann06aa} termed \textit{predictability of the third kind} our ability to anticipate bifurcation phenomena, by reference to the predictabilities of the \textit{first} and \textit{second} kind originally introduced by Lorenz \citep{Lorenz75aa}.

The article is structured as follows. 
Section 2 reviews some of the basic concepts of oscillator theory. This is no substitute for proper textbook reading, but the reader will find essential notions and definitions needed to understand the remainder. Section 3 reviews how these concepts enter theories of ice ages and rapid events. Section 4  discusses effects of stochastic fluctuations and, finally, 
section 5 is a more personal statement about the potential for inference with simple stochastic dynamical systems in palaeoclimate science. 



\begin{figure}[h!]
\begin{center}
\includegraphics[width=0.95\textwidth]{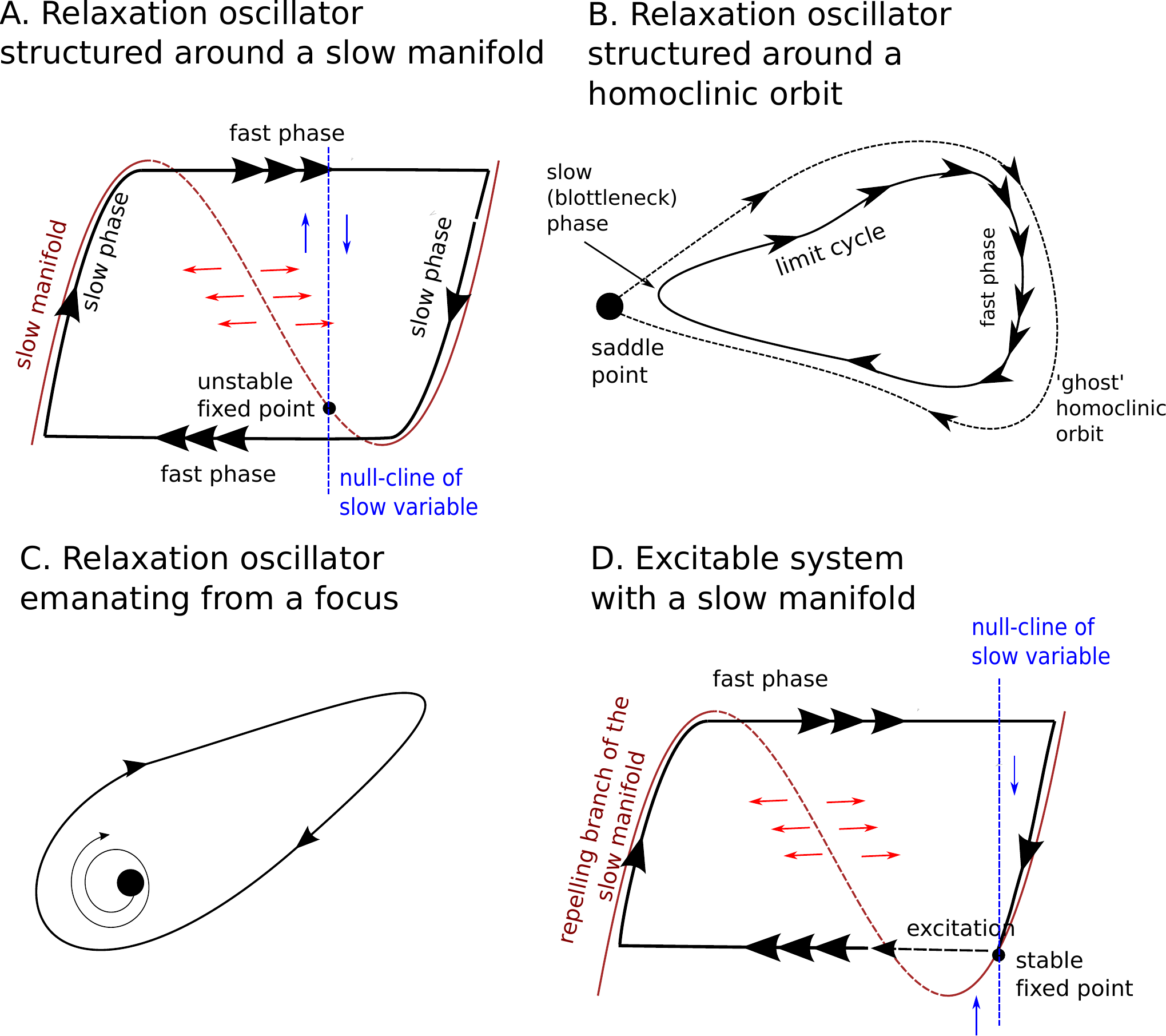}
\end{center}
\caption{
Sketch of  several forms of relaxation oscillations. 
\emph{(A.)} The vector space is structured by a slow manifold with several stable branches. All points of the state space are attracted towards the stable branches of the slow manifold (in full line) along the fast direction, which is here the horizontal. The slow evolution consists in an upward or downward course along the slow manifold depending on whether the system lives on the right- or left-hand side of the null-cline of the slow-variable. The relaxation oscillation consists in alternate jumps between the two branches of the slow manifold. 
\emph{(B.)} Trajectories are rapidly attracted towards a region of the phase-space influenced by a saddle point.
In the scenario dispayed here, there exists a combination of parameters for which the limit cycle crosses the saddle point. In that case, the period of the orbit, which includes the saddle-node, is infinitely long. It is a `homoclinic orbit', hence this particular bifurcation is named a  homoclinic bifurcation. The homoclinic orbit only exists at the bifurcation point, but it influences the orbit when the parameter is close to the bifurcation. This is the reason why one refers to the `ghost' of the homoclinic orbit. There is another scenario for which different saddle-nodes are connected to each other. The orbit and the associated bifurcation are then said to be 'heteroclinic'. 
\emph{(C.)} The relaxation oscillation is organised around a fixed point, with complex eigenvalues with a positive real part. The bifurcation giving rise to this orbit is a Hopf bifurcation.
\emph{(D.)} One example of excitability, here depicted for a slow-fast system. The system resides in a stable space, but a fluctuation may cause an ejection out of the unstable (dashed) branch of the slow manifold. The system then loops all the way through the slow manifold before coming back at rest. 
}
\bigskip
\label{fig:sketch_oscillation}
\end{figure}

\section{Vocabulary and elementary notions.}
The  reader will find an accessible introduction to dynamical system theory and concepts in ref. \citep{Strogatz94aa}. More formal background on oscillator theory, albeit a bit dated, is available in ref.  \citep{Guckenheimer83aa}. Bifurcation and oscillator theory is explicitly connected to climate theory in ref. \cite{Ghil87aa} (see, in particular, chapter 12) and ref. \cite{saltzman02book}, chapter 7.  Background on synchronisation and an introduction to the phenomenon of excitability is available in ref. \citep{Pikovski01aa}. 
Finally, the Scholarpaedia peer-reviewed web-site is an increasingly rich and authoritative source of information on dynamical systems. Only the notions essential for the present article are summarised here. 
\begin{description}
\item[Oscillator:]
The oscillator is a dynamical system that has a globally attracting limit cycle. In more simple terms, it oscillates even in absence of an external drive. Here, we are interested in oscillators to describe climate phenomena, which involve dissipation of energy. The minimal model for a  \textit{dissipative} oscillator includes two ordinary differential equations, of which at least one is non-linear. 
\item[Relaxation oscillator:] 
The relaxation oscillator is a particular kind of oscillator featuring an interplay between relaxation dynamics (generally fast) and a destabilisation process (generally slow). The relaxation is the process by which the system is attracted to a region of the phase space. This evokes the relaxation of a spring. In a relaxation oscillator the system continues to evolve slowly after the relaxation phase. During this slow evolution phase the system stability diminishes gradually until the system is ejected out of its relaxation state, either towards another relaxation state, or to the same relaxation state via a dissipative loop. 
In this review we will encounter three kinds of relaxation oscillators \figrefp{fig:sketch_oscillation}: 
relaxation founded on slow-fast dynamics (involing a slow manifold); relaxations structured by a homoclinic orbit (involving only one relaxation state), and relaxations structured around a focus. More details are given in the caption of \figref{fig:sketch_oscillation}.
\item[Excitability:] An excitable system has a globally attracting fixed point (it does not oscillate spontaneously). However, an external perturbation may have the effect of \textit{exciting} it. During this excitation, the system is being ejected far from its fixed point and then  returns to it. 
\item[Link between relaxation dynamics and excitability:] In practice it is often found that a relaxation oscillator may be transformed into an excitable system by a mere change in parameter, and vice-versa. 
The reason is the following. A relaxation oscillation is often structured globally in the phase space, for example by a slow manifold (\figref{fig:sketch_oscillation}A) or by one or several saddle points \figref{fig:sketch_oscillation}B). 
Suppose now that the oscillation displayed by such a system ceases because a parameter has been changed. The system is then no longer an oscillator, but the `backbone' of the oscillation dynamics are still latent in the phase space because 
the elements that structured the limit cycle (the slow manifold or the saddle points) have not disappeared. 
Consequently, the system may be run on a trajectory close to the defunct limit cycle if it is being pushed by some external force  (the excitation) into the region of the phase space previously occupied by this limit cycle.
This point is illustrated on the basis of slow-fast dynamics on \figref{fig:sketch_oscillation}D, but similar excitation dynamics generally occur near any kind of `explosive bifurcation', that is, bifurcations that give rise rapidly to a fully developed limit cycle. This includes homoclinic, heteroclinic, and certain Hopf bifurcations (two examples follow and are illustrated on \figref{fig:bifpp4}). 
\end{description}

\section{Oscillators, relaxation and excitability in palaeoclimates \label{sec:paleomodels}}
\subsection{Models of ice ages}
\subsubsection{The Saltzman et al.  models.} Saltzman established a theory in which ice ages are interpreted as a limit cycle synchronised on the astronomical forcing. Saltzman and his collaborators wrote a series of articles on the subject, starting with the introduction of the limit cycle idea \citep{Saltzman84ab} and synchronisation hypothesis \citep{saltzman84aa}, the interpretation of the Middle Pleistocene Transition as a bifurcation \citep{saltzman87},  and the more complete models in the mid-1990s \citep[e.g. ref. ][]{saltzman93}. The full theory is developed in a book \citep{saltzman02book}. Here, we concentrate  on two intermediate models \cite{saltzman90sm,Saltzman91sm}. They are called SM90 and SM91, by reference to the authors (Saltzman and Maasch) and the year of publication. The variables $I$, $\mu$ and $\theta$ are the continental ice mass, CO$_2$ concentration and deep-ocean temperature, respectively. 
The reader is referred to the original publications for the meaning and value of the different parameters.  They are not crucial here; it suffices to know that they are all positive. 
\begin{equation}
\left\{
\begin{split}
\ddt{I}&=  \alpha_1 - (c\alpha_2)\mu - \alpha_3 I - k_\theta \alpha_2 \theta - k_R \alpha_2 F_I(t)  \\ 
\ddt{\mu}&= \beta_1 - (\beta_2 - \beta_3 \theta + \beta_4 \theta^2)\mu - (\beta_5 - \beta_6 \theta)\theta + F_\mu (t)\\ 
\ddt{\theta}&=  \gamma_1 - \gamma_2 I - \gamma_3 \theta 
\label{sm90}
\end{split}
\right.
\tag{SM90}
\end{equation}
and 
\begin{equation}
\left\{
\begin{split}
\ddt{I}&=  \alpha_1 - (c\alpha_2)\mu - \alpha_3 I - k_\theta \alpha_2 \theta - k_R \alpha_2 F_I(t) \\ 
\ddt{\mu}&= \beta_1 - (\beta_2 - \beta_3 \mu + \beta_4 \mu^2)\mu - \beta_5 \theta + F_\mu(t) \\ 
\ddt{\theta}&=  \gamma_1 - \gamma_2 I - \gamma_3 \theta 
\label{sm91}
\end{split}
\right.
\tag{SM91}
\end{equation}

In both  models the first equation describes the ice mass response to changes in CO$_2$ ($\mu$) and the astronomical forcing ($F_I(t)$) 
Saltzman adopts the so-called Milankovitch view \footnote{In fact, this view is introduced by Murphy \citep{Murphy76aa} but it is developed mathematically in the `Canon of Insolation' \citep{milankovitch41} authored by Milankovitch} that an increase in insolation causes a decrease in ice mass.  Increases in CO$_2$ or in ocean temperature have the same effects. 

The other two equations describe the dynamics of CO$_2$ and the response of deep-ocean temperature to changes in ice volume. 
It is further assumed that the mean state of climate varied slowly  throughout the Pliocene-Pleistocene, in particular in response to a `tectonically-driven' decline in the average concentration in CO$_2$, consistently with an earlier proposal \citep{Raymo88aa}. This tectonically-driven decline is here modelled as a slow decrease in the forcing term $F_\mu(t)$ throughout the Pleistocene.

\begin{figure}[t!]
\begin{center}
\includegraphics[width=0.45\textwidth]{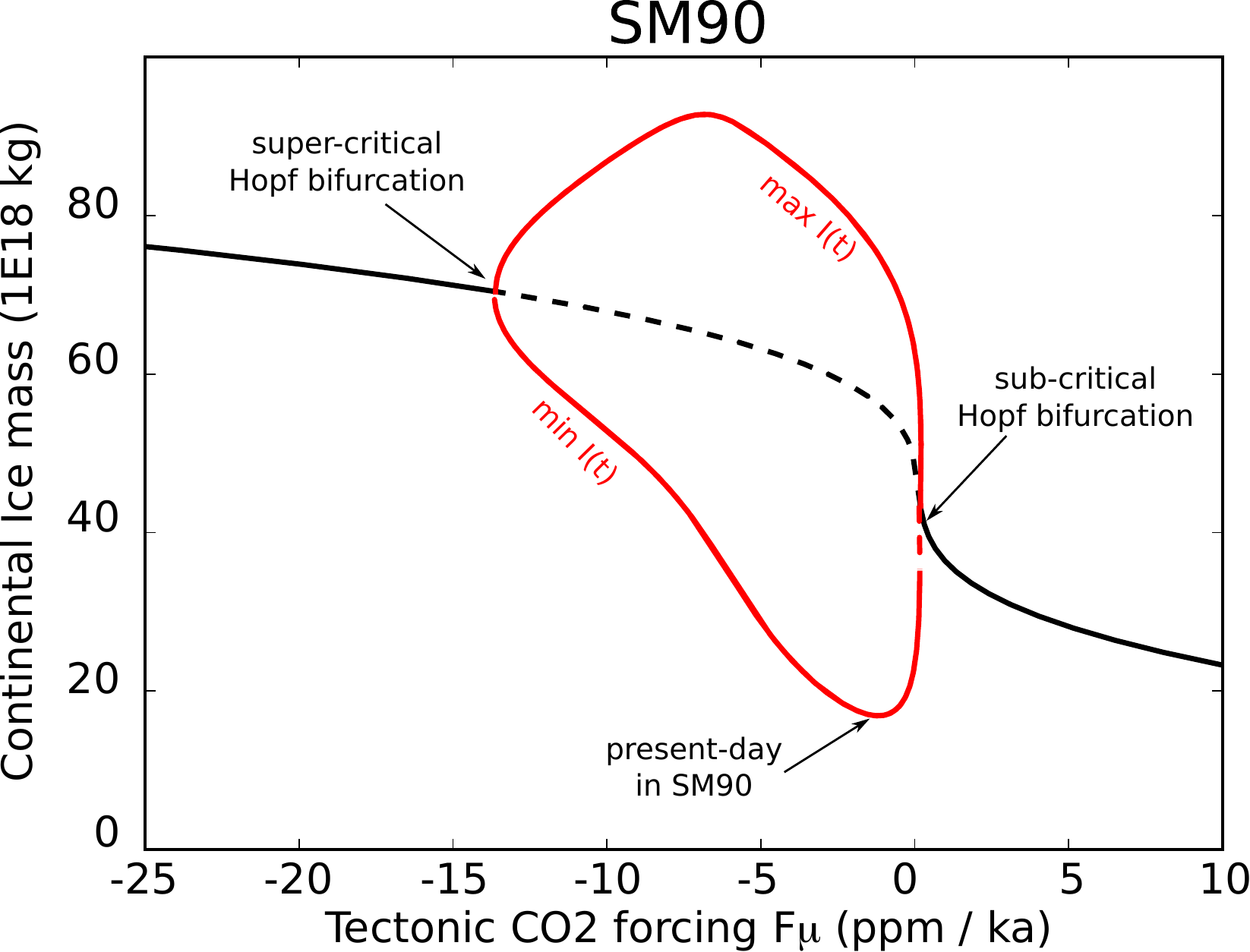}
\includegraphics[width=0.45\textwidth]{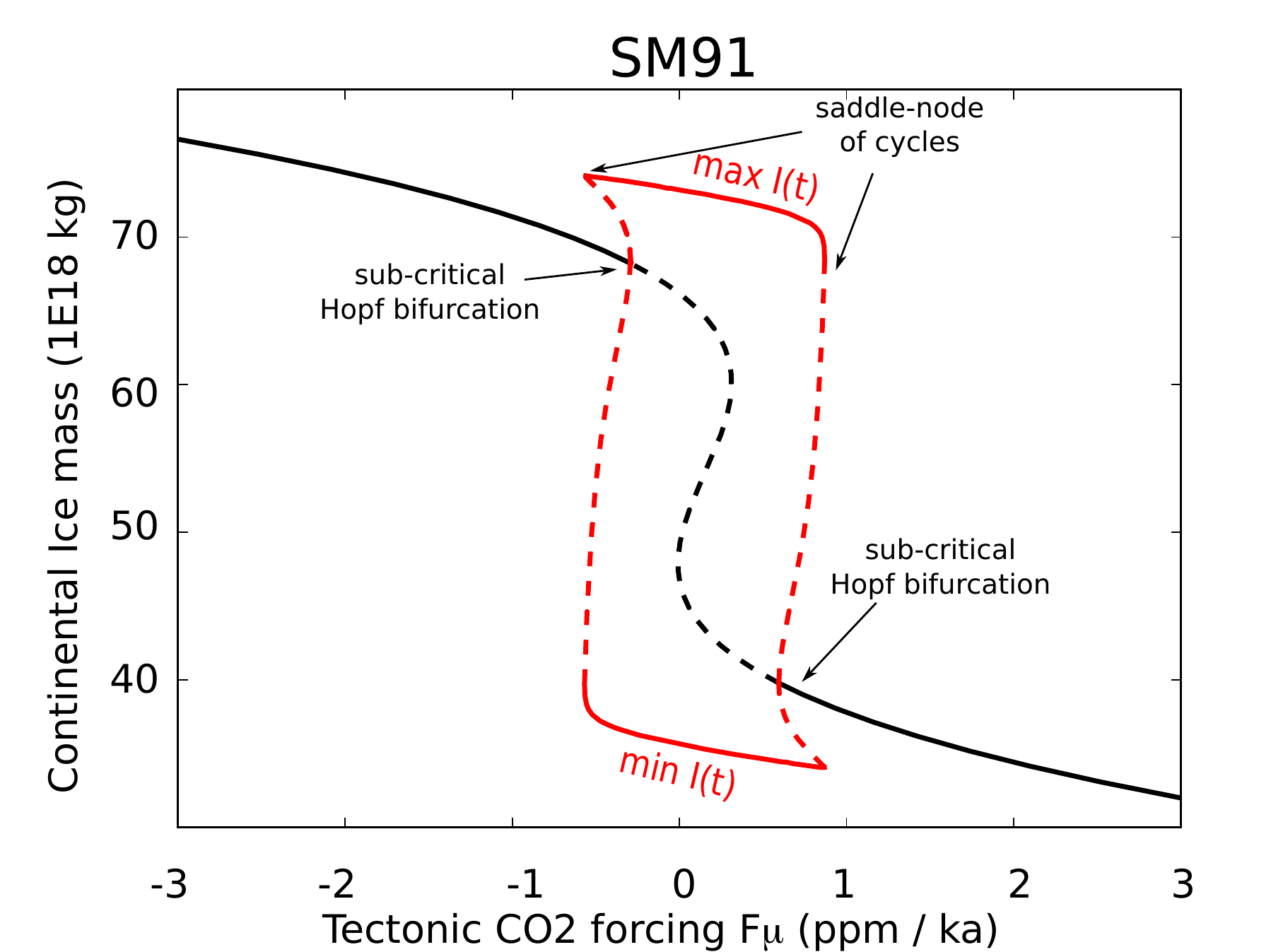}
\end{center}
\caption{Bifurcation diagrams of the Saltzman and Maasch models (SM90) and (SM91) as a function of $F_\mu$, here treated as a constant control parameter. Note the difference in scales on both axes. Black lines are fixed points. Continental ice mass $(I)$ is shown as a function of tectonic forcing $F_\mu$.  The red lines indicate  limit cycles, shown as the minimum and maximum values of $I$ along the limit cycle.  Unstable fixed points or limit cycles are denoted by dashed lines. Calculations and figures were made using the pseudo arc-length continuation software package AUTO \citep{auto07p}.}
\label{fig:bifsm}

\begin{center}
\includegraphics[width=0.45\textwidth]{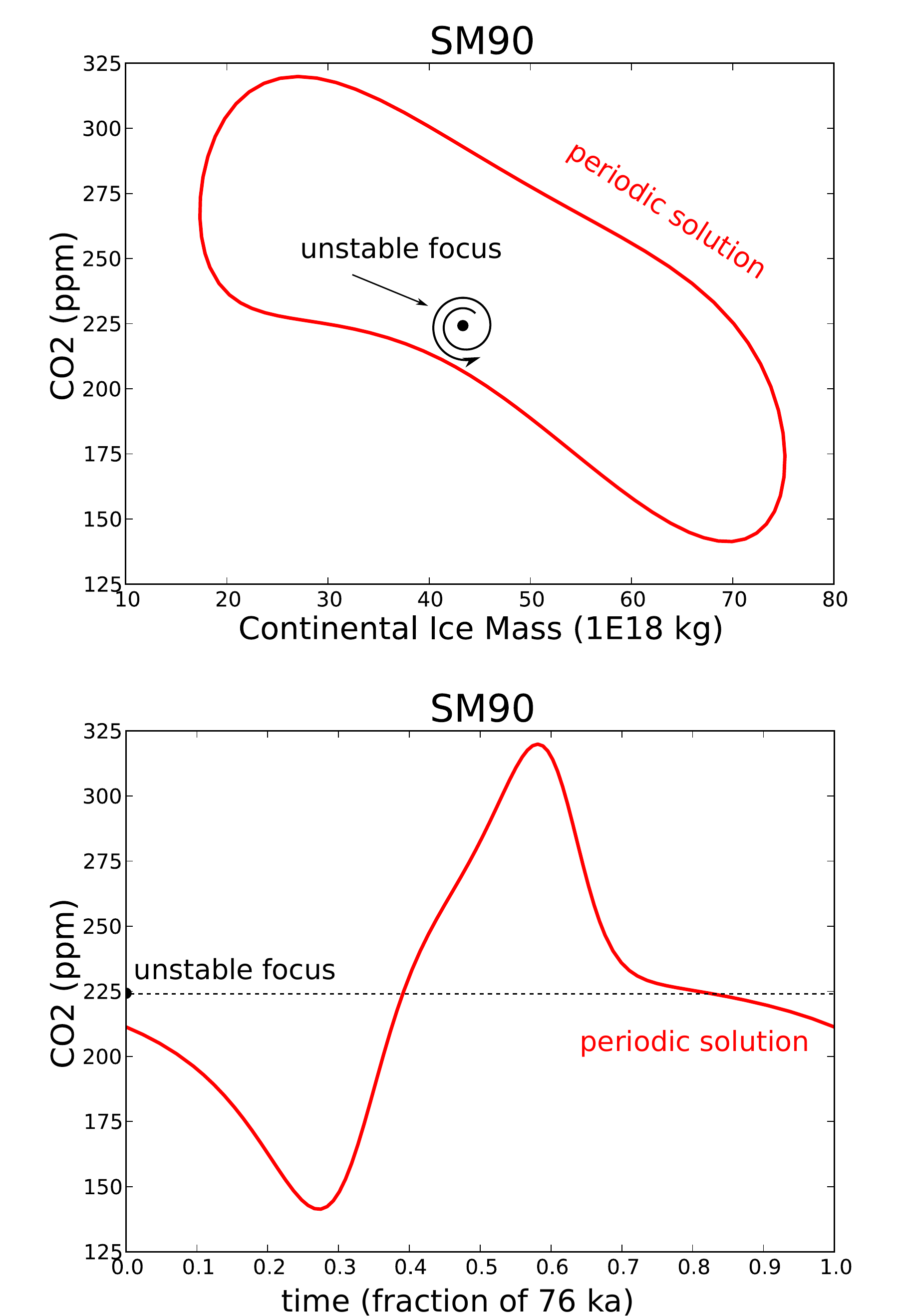}
\includegraphics[width=0.45\textwidth]{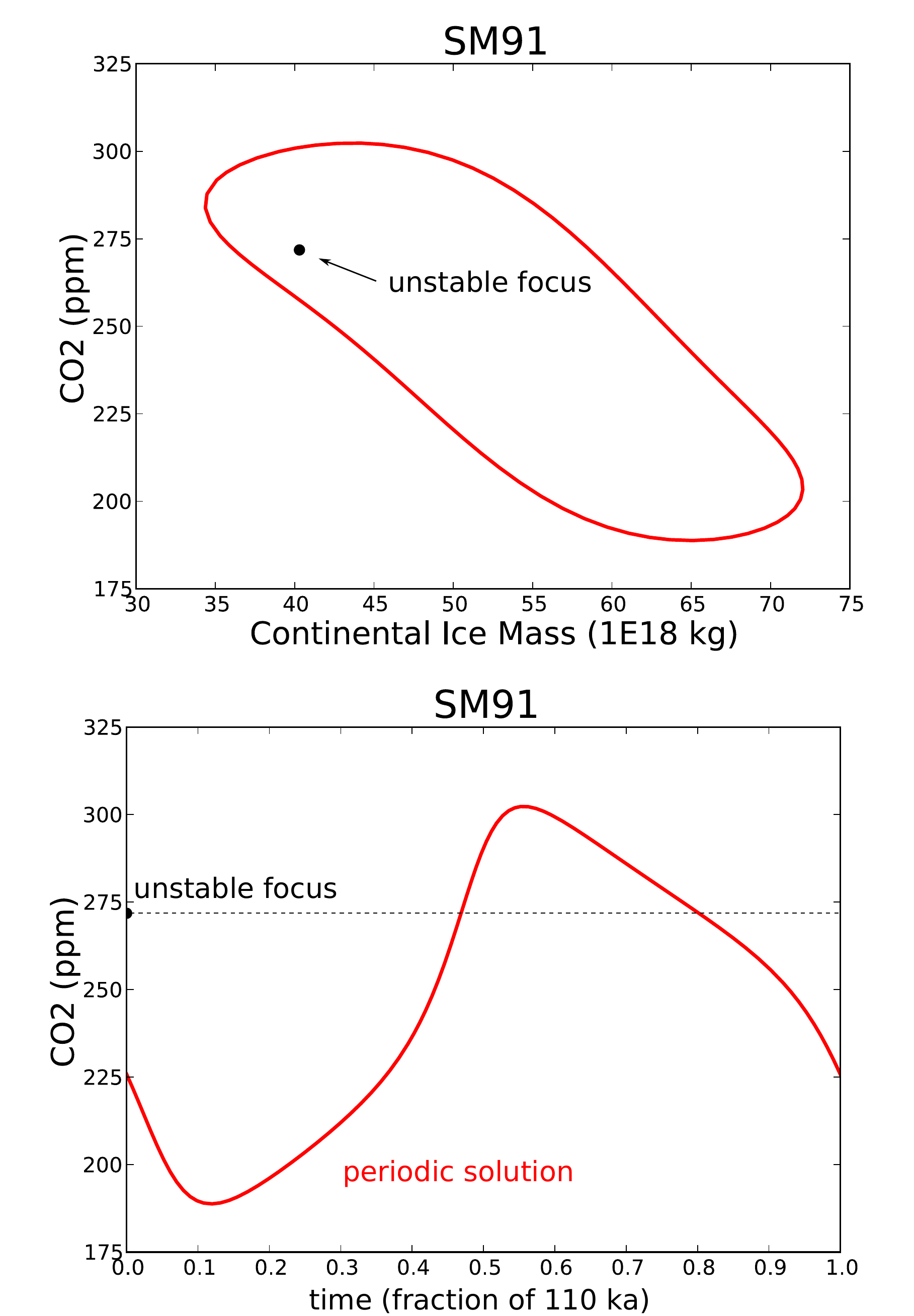}
\end{center}
\caption{Periodic and fixed point solutions for SM90 (with $F_\mu = 1.2$ka/ppm) and SM91 (with $F_\mu = 0.5$ka/ppm), near the sub-critical Hopf bifurcation points. The dynamics of SM90 slow down near the unstable fixed point, while the limit cycle of SM91 is much more decoupled from the position of the fixed point. }
\label{fig:solsm}
\end{figure}
\begin{figure}
\begin{center}
\includegraphics[width=\textwidth]{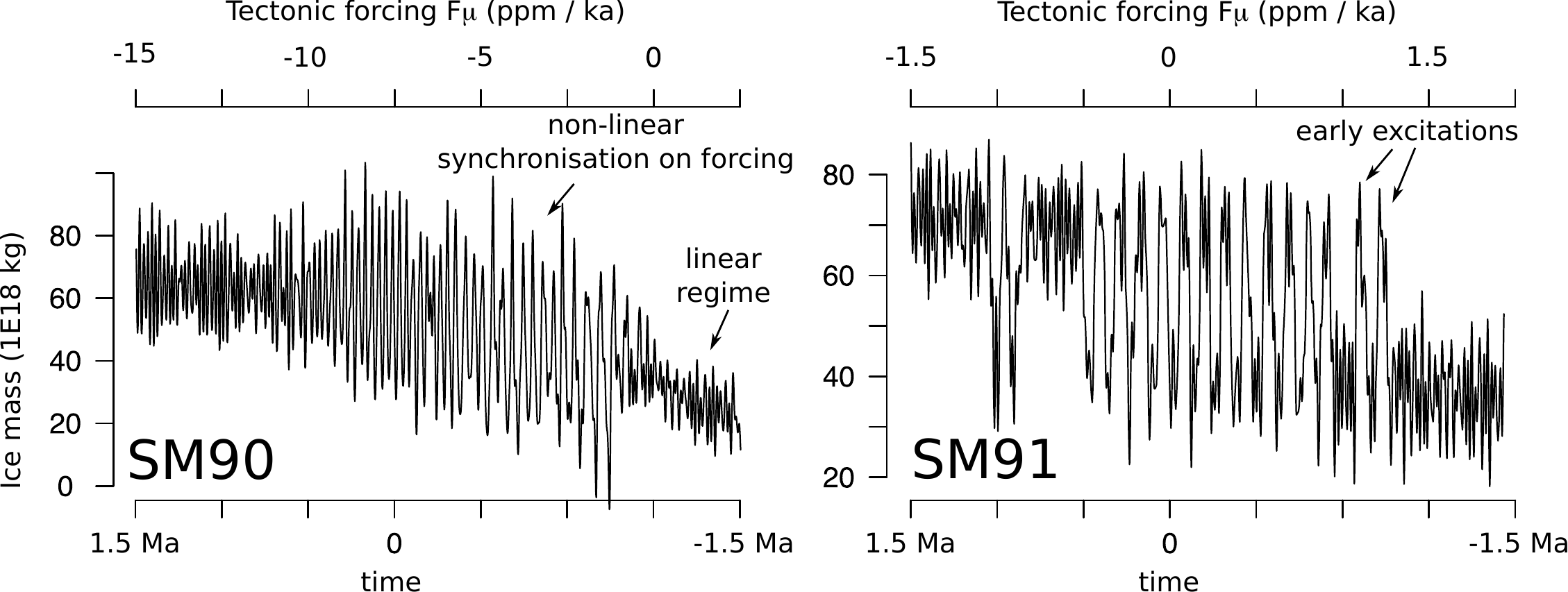}
\end{center}
\caption{Two histories of ice volume generated with the same models (SM90) and (SM91), using the astronomical forcing and a decline scenario for the tectonic forcing of CO$_2$. The scenarios were chosen here to evidence the rise and decline of 100,000-year ice ages and are not the same as those used by Saltzman. Observe the explosive character of the appearance and decline of 100,000-year cycles in SM91, with early and late excitations of the limit cycle by the astronomical forcing, and the smoother character of the evolution on SM90. Time is running from right to left. }
\label{fig:hopf_saltzman}
\end{figure}

Consider the bifurcation diagrams of the  SM90 and SM91 models with respect to $F_\mu$, assuming no astronomical forcing ($F_I=0$)  \figrefp{fig:bifsm}. The systems are then said to be free or autonomous. 
Depending on $F_\mu$, both models show regimes with a stable fixed point, and regimes for which the fixed point is unstable so that the system orbits along a limit cycle. 

These considerations led Saltzman to interpret the Middle Pleistocene Transition as a bifurcation between a `quasi-linear' response regime to the astronomical forcing (in the fixed-point regime) to a regime of non-linear synchronisation (resonance) on the astronomical forcing.
He concluded that ice ages would occur today even in absence of astronomical forcing.  The main effect of the astronomical forcing is to control the timing of glaciations.

Saltzman's theory is seductive because it explains in a consistent framework several aspects of the Pleistocene climate history, including the change from linear to non-linear regime   \citep{lisiecki07trends}, the presence of 100,000 year periodicity in climate records  \citep{hays76}, the lack of  a 400,000-year spectral peak in the ice-volume record (such a peak appears in the simple piece-wise linear model devised by Imbrie and Imbrie \citep{imbrie80}, due to rectification of the precession signal), the synchronisation of deglaciations on the astronomical forcing \citep{Raymo97aa,huybers07obl}, and the occurrence of large climatic transitions even when eccentricity, which modulates the effect of precession on insolation, is at its lowest.

The difficulty for accepting Saltzman's models as a definitive theory lies in the physical interpretation of the CO$_2$ equation. This equation  encapsulates  all the interesting dynamics of the system and it is thus crucial to the theory. 
Some semi-empirical justification for the CO$_2$ equation is given in ref.  \citep{saltzman87} but the form of this equation  has undergone some somewhat ad hoc adjustments in SM90. The form present in SM91 is again different, with important effects on the bifurcation structure, while the authors did not justify this latter change based on physical or biegeochemical considerations. 

To better appreciate the stuctural differences between the two models let us return to the bifurcation diagrams.
Consider SM90. As  the forcing is decreased the fixed point gives rise to a locally unstable limit cycle. The system must therefore find a stable limit cycle further away from the fixed point, but in this case not much further. 
This stable limit cycle is under the influence of the unstable fixed point, and in particular the system slows down when it passes near it \figrefp{fig:solsm}. This is  scenario `C' depicted on \figref{fig:sketch_oscillation}. The limit cycle evolves as the tectonic forcing is further decreased, until it shrinks smoothly around a perpetually glaciated state. 

In SM91, the bifurcation induced by the decrease in tectonic forcing is much more explosive. The system lands on a stable limit cycle that turns out to be little affected by the position of the unstable point. The cycle dynamics do not show clear phases of acceleration and the system cannot be regarded as a relaxation system. The limit cycle disappears abruptly as the tectonic forcing is further decreased, through a phenomenon called a saddle-bifurcation of cycles. The consequences of the difference between the bifurcation structures of SM90 and SM91 may be further appreciated in the transient experiments shown on \figref{fig:hopf_saltzman}.

\subsubsection{Paillard's (1998) ice age model (P98).}
Paillard has been advocating the concept of relaxation for understanding palaeoclimate dynamics, both  ice ages and the more abrupt events, since the publication of a seminal paper \citep{Paillard94aa} in 1994. 
We return to this article later on, and concentrate on another article published in 1998 \cite{paillard98}, in which Paillard introduces a conceptual model of ice ages. 
Ice volume dynamics respond to an ordinary differential equation:
\begin{equation}
\frac{\mathrm d x }{\mathrm d t} = \frac{x_R(y) - x}{\tau_R(y)} - \frac{ F(t)}{\tau_f} \tag{P98}
\label{}
\end{equation}
In this equation, the ice volume $x$ is linearly \textit{relaxed} to  $x_R$ with characteristic relaxation time $\tau_R$. This relaxation process is further perturbed by the astronomical forcing $F(t)$ with a characteristic time $\tau_f$. 
Such a system is said to be hybrid \citep{Guckenheimer03aa} because the relaxation equation involves a discrete state variable, here denoted $y$. Its state  may be `deep glacial' ($G$), `mild glacial' ($g$) or `interglacial' ($i$). The numerical values of $x_R$ and $\tau_R$ depend on this climate state. Climate states $y$ follow a sequence $i\rightarrow g \rightarrow G$ according to a set of conditions formulated on the level of glaciation $x$ and insolation. Namely, the transition $g\rightarrow G$ is triggered when the forcing $F(t)$ exceeds a certain threshold. Occurrence of $G$ drives climate quickly into an interglacial state $i$ because $x_R(G)$ and $\tau_R(G)$ are specified in the model to be low. 

Paillard is not very specific about the physical meaning of the discrete variable, but it accommodates the paradigm that the Atlantic ocean circulation has gone through three different states during the latest glacial period: intermediate circulation, shut-down of the circulation, and modern, deep-sinking circulation. 
%
The system (P98) features the concept of slow-fast relaxation dynamics. 
However,  this is \textit{not} an oscillator because the shift from $g$ to $G$ is determined by the course of the external forcing. 
The Middle Pleistocene Transition is induced in (P98) in a fashion similar to Saltzman, and on the basis of similar physical assumptions (tectonically-driven decline in CO$_2$). The drift in climatic conditions induced by tectonics is accounted for by a term added to the astronomical forcing. 
In a later review, Paillard \citep{paillard01rge} further emphasises empirical evidence  for the relevance of the relaxation concept in the phenomenon of deglaciation.
\subsubsection{The Gildor Tziperman model.}
Gildor and Tziperman \citep{Gildor-Tziperman-2000:sea}  take a moderate step towards higher model complexity by considering a slightly more explicit representation of atmosphere, ocean, sea-ice and land-ice dynamics. Namely, the ocean is divided into 8 boxes, and the atmosphere into 4. Sea-ice fraction responds to standard energy balance equations. More crucially, land-ice growth is influenced by a somewhat controversial feedback between sea-ice and precipitation. The feedback is controversial because it is assumed that cold climate results in a \textit{reduction} in ice volume: sea-ice growth causes a reduction in precipitation in ice-covered areas and, by this mechanism, almost suppresses accumulation of snow on ice sheets. The latter  then no longer compensates for ice ablation and ice volume shrinks. 

A free oscillation arises from the fact that the ice volume thresholds for switching sea-ice cover `on' and `off' differ. In other words, sea-ice displays a hysteresis response to variations in ice volume.  This is exactly the principle of the slow-fast relaxation oscillator depicted on \figref{fig:sketch_oscillation}A : The curve of equilibrium of sea-ice with respect to ice volume is the slow manifold, and ice volume integrates the state of sea-ice in time. In turn, this oscillation can be synchronised on the astronomical forcing.

The Gildor-Tziperman model is coupled to a biogeochemical cycle in a companion paper  \citep{Gildor01aa}, but the essential dynamics of the glacial oscillation are unchanged.  Tziperman et al.  \citep{tziperman06pacing} further  comment on the model and its property of synchronisation on the astronomical forcing, and find that its behaviour is essentially reducible to a hybrid dynamical system. 
\subsubsection{The Paillard-Parrenin model.}
  Paillard and Parrenin \citep{paillard04eps} propose yet another relaxation model in 2004 (PP04). The prognostic variables are ice volume $I$, the area of the Antarctic continental ice sheet $A$ and the atmospheric concentration in CO$_2$ ($\mu$) ($a\ldots j$ are parameters):
\begin{equation}
\left\{
\begin{split}
\ddt I &= \frac{1}{\tau_I}(- a \mu - b F(t) + c - I) \\
\ddt A &= \frac{1}{\tau_A}( I - A) \\
\ddt \mu &= \frac{1}{\tau_\mu}( d F(t) - e I + f H(-D) + g -\mu)\\
D &= h I - i A + j
\end{split}
\right.
\tag{PP04}
\label{PP04}
\end{equation}
  
  As in the other ice age models, ice volume is a slow variable driven by the astronomical forcing. It is here coupled to a variable with a similar time scale ($\tau_A \sim \tau_I$) and a faster one ($\tau_\mu =  \tau_I/3$). The term $H(-D)$, where $H$ is the Heaviside function, represents the ventilation of the Southern ocean. CO$_2$ is  released into the atmosphere when the Southern ocean is ventilated ($D<0$), which drives deglaciation. Ice then grows slowly, until a Southern ocean ventilation flush sends the system back to interglacial conditions. Ocean ventilation is thus the fast process in this model and it is the only non-linear process accounted for. Though, contrary to the Gildor-Tziperman model, it \textit{does not} present a hysteresis behaviour. Consequently, the glacial cycles featured by this model cannot be interpreted in terms of shifts between the branches of a slow manifold. 

\begin{figure}[t!]
\begin{center}
\includegraphics[width=0.45\textwidth]{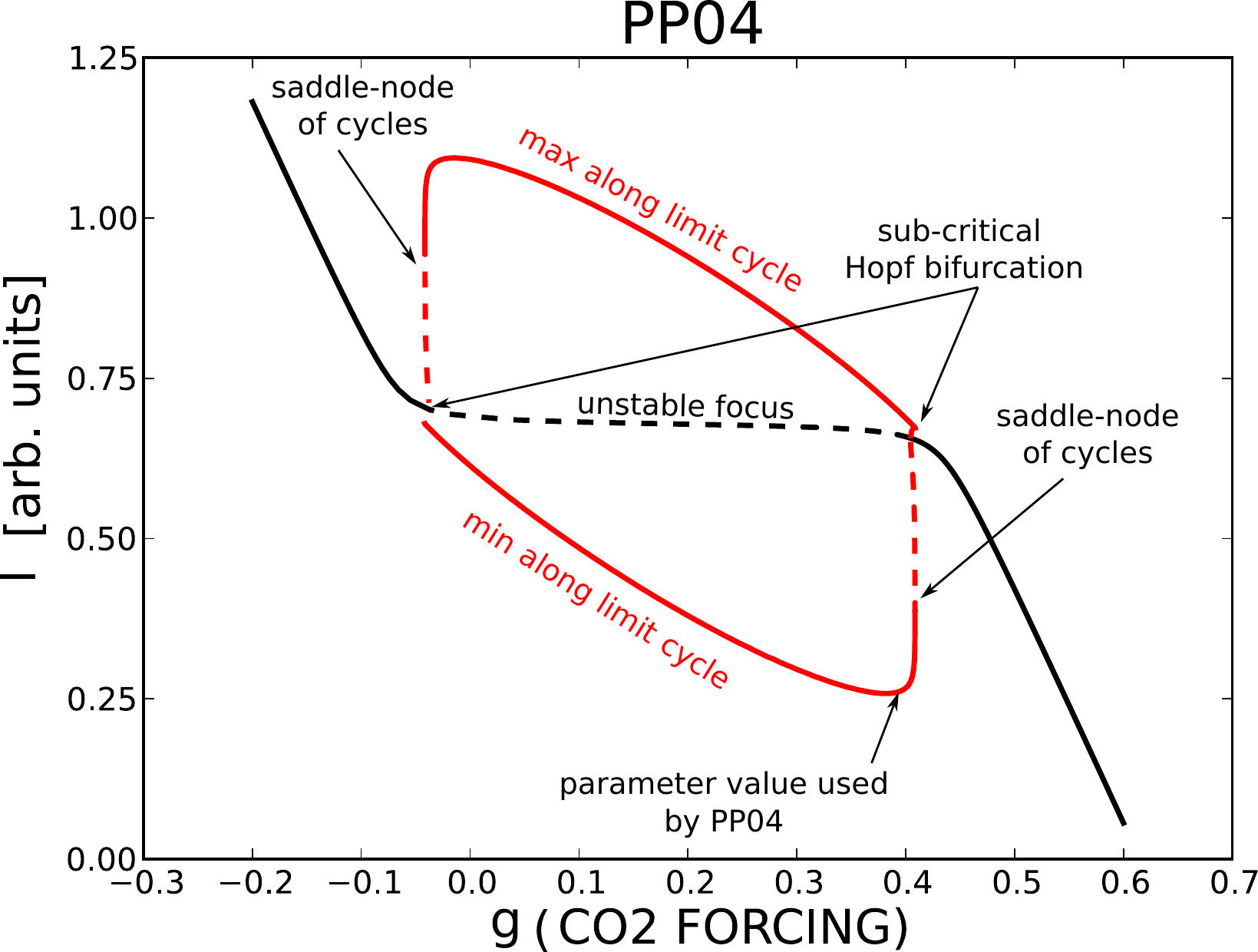}
\includegraphics[width=0.45\textwidth]{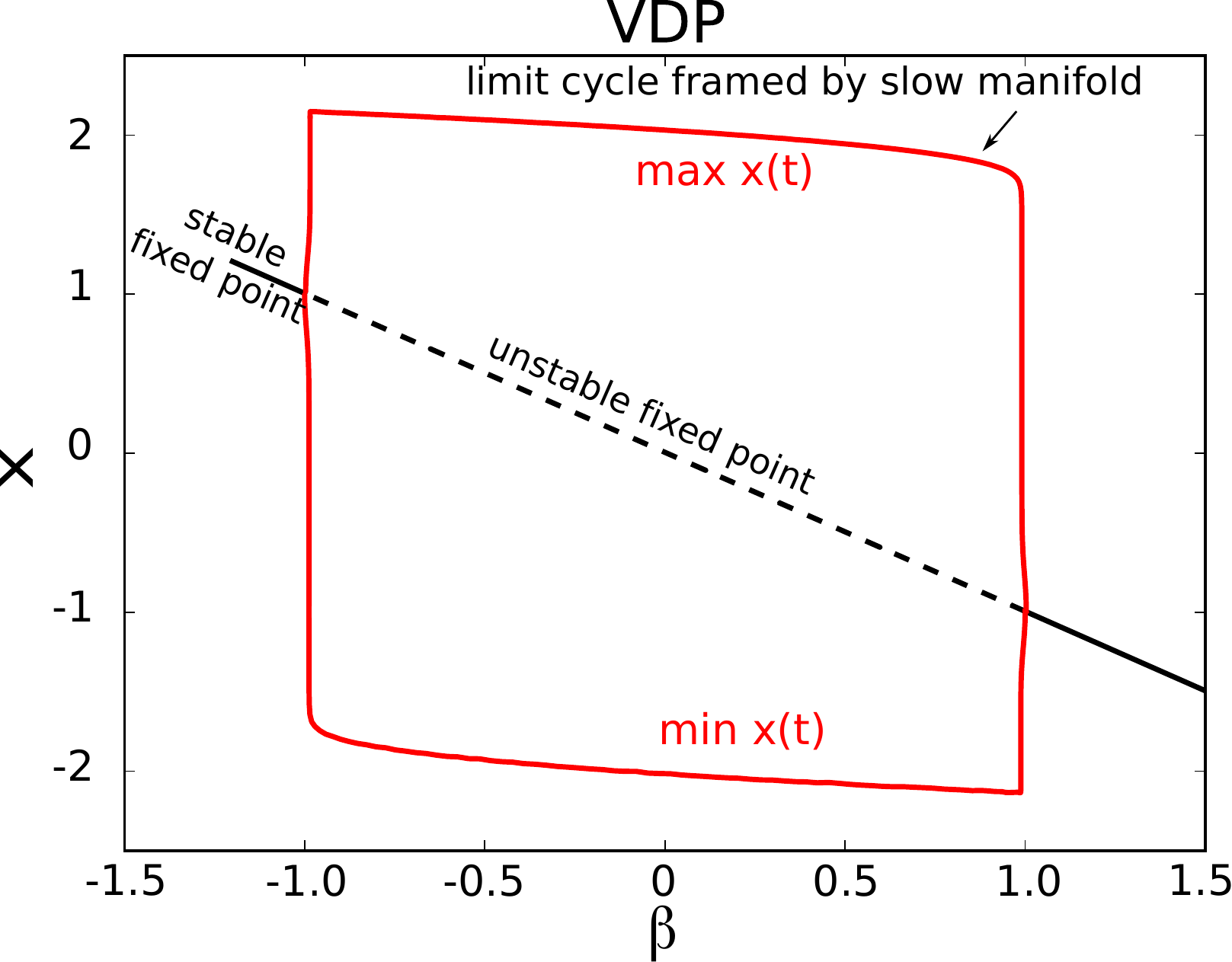}
\includegraphics[width=0.45\textwidth]{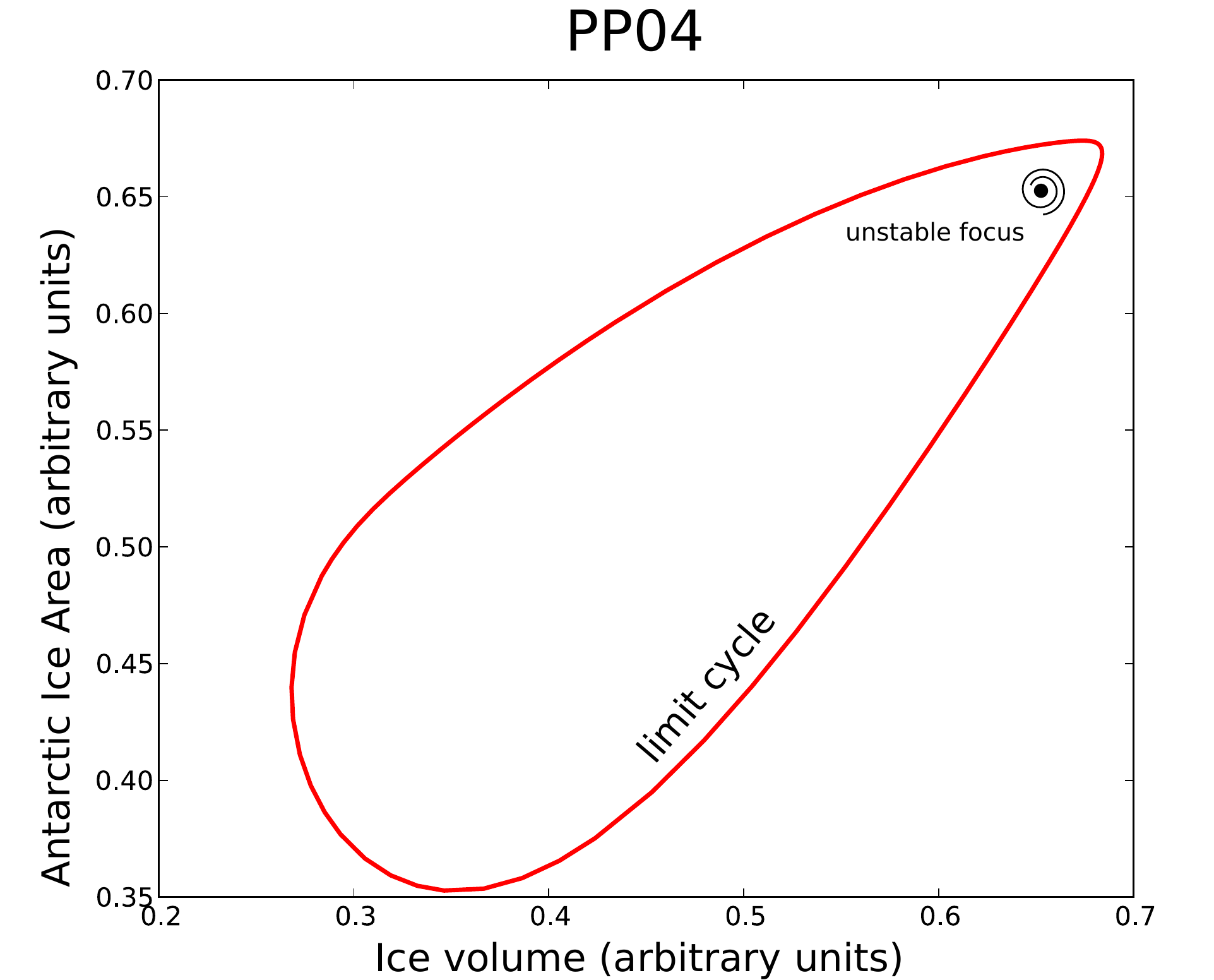}
\includegraphics[width=0.45\textwidth]{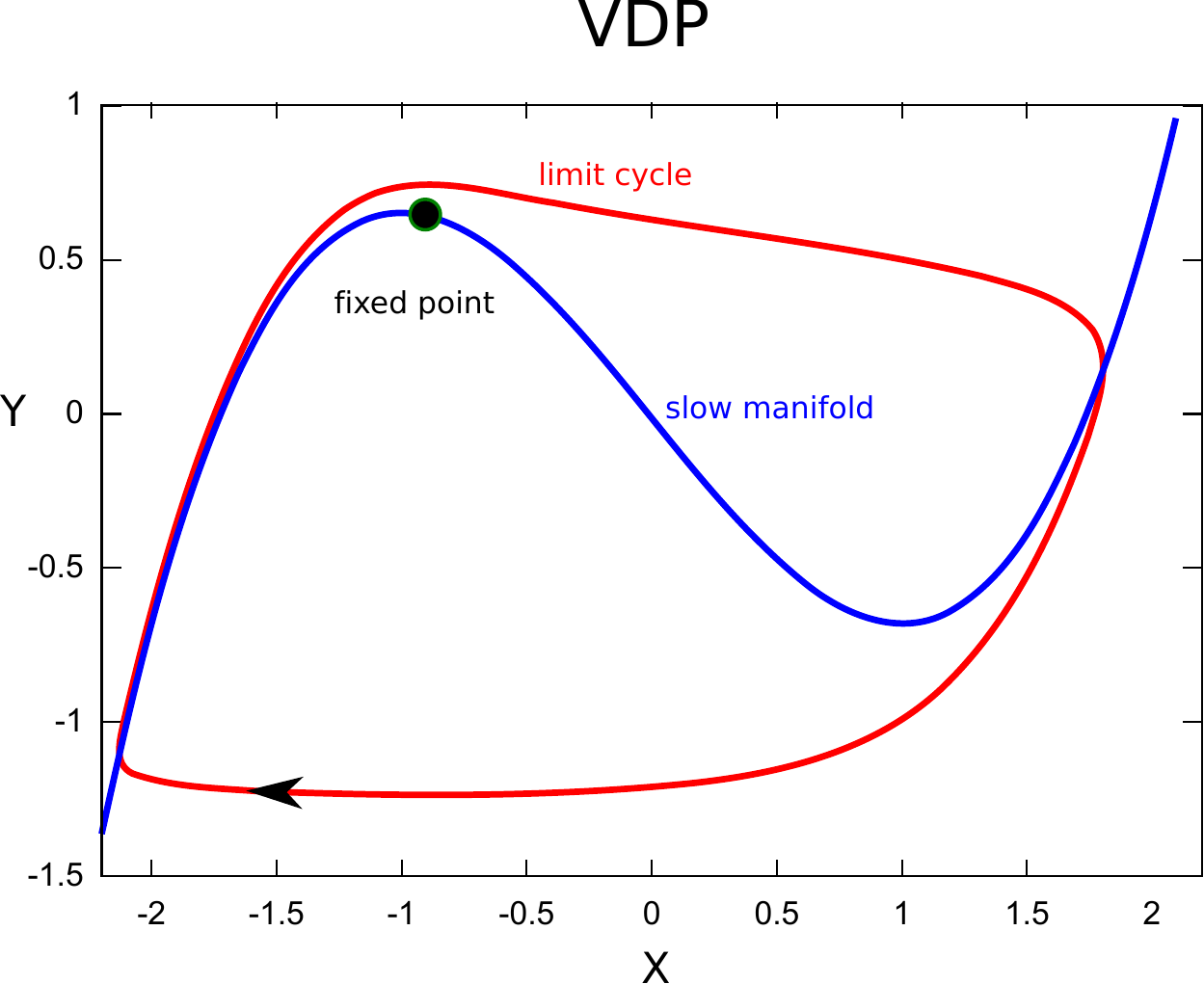}
\end{center}
\caption{Bifurcation diagrams and phase-space trajectories of the free Paillard-Parrenin model (PP04) and the van der Pol (VDP) model. Both systems display explosive bifurcation scenarios, but the details are different. PP04 exhibits a sub-critical Hopf bifurcation, while the limit cycle of VDP is explicitly framed by a slow manifold. 
Phase space trajectories are drawn near the bifurcation points, that is : $g=0.4$ in PP04 and $\beta=0.9$ in VDP.  The Heaviside function in PP04 is approximated as $H(x) = \mathrm{atan} (500 x) / \pi + 0.5 $ for analysis with AUTO.}
\label{fig:bifpp4}
\end{figure}

 To better understand the dynamics of glacial cycles in this model we consider  the bifurcation diagram along typical solutions in the phase space for the free (i.e. unforced) system \figrefp{fig:bifpp4}. The parameter $g$ is taken in this example as the control parameter, in order to preserve Saltzman's idea that ice age cycles appear as the consequence of a slow perturbation of the carbon cycle. As in SM90, PP04 exhibits a limit cycle arising from a sub-critical Hopf bifurcation. 
  The dynamics along the limit cycle close to the bifurcation point are strongly influenced by the presence of the unstable focus.
  This is the configuration  `C' shown in  \figref{fig:sketch_oscillation}C.  Depending on $g$, the focus is either on the low-ice-volume side of the limit cycle (i.e.: the system spends most of its time with high CO$_2$) or on the high-volume side of the limit cycle (i.e.: the system spends most of its time in low CO$_2$). Parrenin and Paillard estimate that we are currently in the second configuration. 

\subsubsection{A minimal model of ice ages.}
It has been claimed \cite{tziperman06pacing,Cane:2006} that any model that has some form of 100,000 year internal periodicity could be used to reproduce the course of ice volume over the last 800,000 years. 
Taking the argument at face value, Crucifix \cite{Crucifix11aa} used one of the simplest possible slow-fast oscillators: the van der Pol oscillator, with minimal modifications to account for the astronomical forcing and the asymmetry between the phase of ice build-up and melt during the late Pleistocene ($\alpha$, $\beta$, $\gamma$, and $\tau$ are parameters; $F(t)$ is the astronomical forcing):
\begin{equation*}
\left\{
\begin{split}
\ddt x &= (- y + \beta + \gamma F(t) ) / \tau\\
\ddt y &= - \alpha (  y^3 / 3 - y - x) / \tau  
\end{split}
\right.
\tag{VDP}
\label{eq:vdp}
\end{equation*}

\begin{figure}[h!]
\begin{center}
\includegraphics[width=\textwidth]{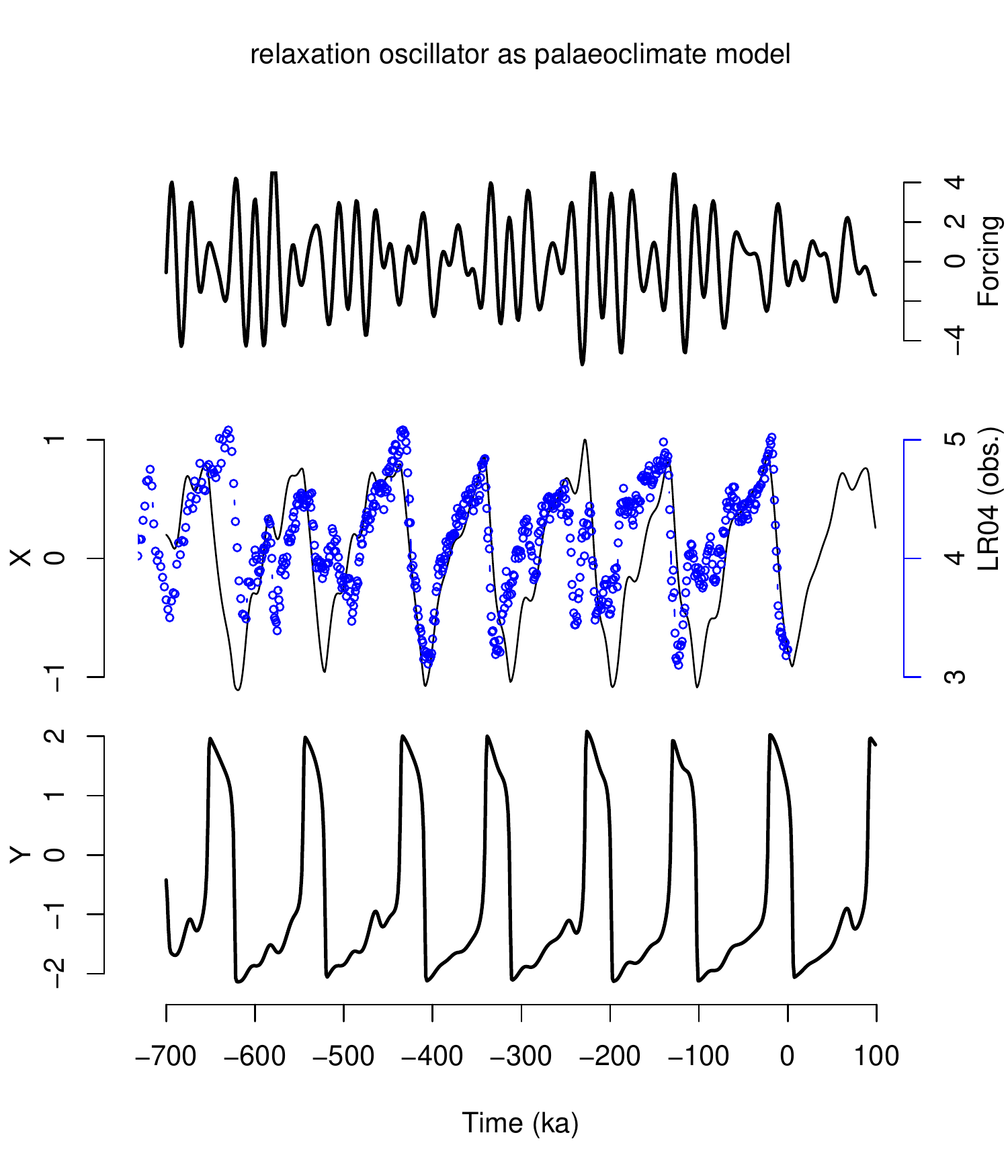} 
\end{center}
\caption{Astronomical forcing, $x$ and $y$ trajectories obtained using system \eqref{eq:vdp} with $\alpha=30$, $\beta=0.75$, $\gamma=0.4$ and $\tau=36\,\mathrm{ka}$ (1~ka = 1,000 years). Blue dots are an authoritative natural archive thought to mainly represent fluctuations in ice volume and deep ocean temperature, and compiled in ref. \citep{lisiecki05lr04}. } 
\label{fig:FitData}
\end{figure}

The system dynamics are determined by the structure of the slow manifold $x = y^3/3-y$. The parameter $\beta$ controls the position of the fixed-point on the slow manifold and, consequently, the ratio of times spent by the system in the two branches (`glacial' and `interglacial') of the slow manifold.  
The ice age curve can be captured with some tuning \figrefp{fig:FitData}, although it is fair to add that 
a small change in parameters may shift the timing of one or several ice age cycles. This minimal model was used 
to challenge intuitive arguments about the predictability of ice ages \cite{Crucifix11aa}. 

\subsection{Models for millennial climate variability}
\subsubsection{Dansgaard-Oeschger events as relaxation oscillations.}
Welander \citep{welander82conv} introduced the concept of relaxation oscillations in the context of ocean dynamics. He described a \textit{heat-salt oscillator}  involving exchanges of heat and salt within a single oceanic column, coupled to a phenomenon of surface temperature relaxation. The destabilisation process needed for the relaxation oscillation to appear is here related to diffusion between the deep ocean and the mixed layer. The system dynamics are further controlled by the mean freshwater flux at the top of the ocean column. It determines the transitions from a regime characterised by perpetual convection in the oceanic column,  to a regime with intermittent convection (oscillation), and finally  to a regime with no convection \cite{Cessi96aa}. The bifurcations between the different regimes bear the character of \textit{global} bifurcations, with the  oscillation period approaching infinity near the bifurcation points (in particular, the second bifurcation bears the character of a homoclinic bifurcation). 
The heat-salt oscillator belongs  thus to the class `B' on \figref{fig:sketch_oscillation}.

Welander \citep{welander86thc} and Winton and Sarachik \citep{Winton93aa} later introduce the concept of another kind of relaxation oscillator in the ocean. It involves the meridional structure of the ocean thermohaline circulation, and the key non-linear process is the meridional advection of heat and salt. The oscillations featured by this model are termed `deep-decoupling' oscillations \cite{Winton93aa}. Given that the slow process  now relates to heat accumulating in the global ocean, the characteristic time of deep-decoupling oscillations is of the order of 1,000 years. The net flux of freshwater delivered to  the North Atlantic acts as a bifurcation parameter controlling the transition between non-oscillating and oscillating regimes in the Winton-Sarachik model \cite{Schulz2002oscillators}. 

Millennial oscillations have since been observed across a hierarchy of ocean models, including  3-D ocean models with prescribed freshwater flux and restoring conditions to surface temperature \citep{Weaver94aa}, and 3-D models coupled to a simple atmosphere \citep{Haarsma01aa, Meissner08aa}.
Sakai and Peltier \cite{Sakai97aa} proposed that millennial deep-decoupling oscillations could explain Dansgaard-Oeschger events. 
 Colin de Verdiere et al. \cite{Colin-de-Verdiere07aa, Colin-de-Verdiere06ab, Colin-de-Verdiere10aa} complement this early proposal with a fairly complete theory based on ocean circulation model experiments. The oscillations described by Colin de Verdiere et al. involve the processes of turbulent vertical mixing (neglected in Winton and Sarachik \citep{Winton93aa}), advection, and convection, which unify the salt oscillator with the deep-decoupling oscillation model. Incidentally, Colin de Verdiere \cite{Colin-de-Verdiere06ab} dismisses the non-linearity of the equation of state as the cause of the oscillations.

There is, across the model hierarchy, consistency about the fact that the transition between the oscillating circulation regime and the so-called diffusive, haline regime (without deep convection) is associated  with a homoclinic bifurcation \citep{Colin-de-Verdiere07aa,Timmermann03ab}. The nature of the bifurcation between the convective regime and the oscillation is more model-dependent. Timmermann et al. \cite{Timmermann03ab}, based on experiments with the 8-ocean-box Gildor-Tziperman model, find a Hopf bifurcation; salt-conserving experiments with a 2-D ocean model show a transition towards a finite-period cycle, but of increasing period as the bifurcation is approached; experiments with a more idealised model, formulated as a 2-equation dynamical system, reveal the signature of an infinite-period bifurcation \citep{Colin-de-Verdiere07aa} \footnote{
This particular case was not illustrated on \figref{fig:sketch_oscillation}. There is no saddle point along or near the orbit, but there is a combination of parameters for which a fixed point appears on the limit cycle. Beyond this particular parameter value, this fixed point splits into a saddle and a node. This particular parameter value correpsonds to an `infinite-period' bifurcation. In practice, as long as the limit cycle exists, the trajectory slows down near the point where the saddle-node will appear. Some authors then refer to the influence of the `ghost' of the saddle point \cite{Strogatz94aa}. }.
 The latter implies that Dansgaard-Oeschger events, at the time when they appear soon after the glacial inception process, should be very long but of a similar amplitude as the Dansgaard-Oeschger events coming later in the glacial cycle. This feature is consistent with the Greenland ice core record \figrefp{fig:data}.
 More specifically, the first Dansgaard-Oeschger cycles that appeared at the beginning of the glacial era were characterised by a long `plateau' phase (also called: interstadial) during which the thermohaline circulation was certainly very active \citep{Capron10aa}. In the Colin de Verdiere et al. theory, the plateau phase is the phase of the trajectory influenced by the `ghost of the saddle point' \citep{Colin-de-Verdiere06ab}. 

\subsubsection{Dansgaard-Oeschger cycles as the manifestation of an excitable system.}
Given the explosive  nature of the bifurcations involved in ocean dynamics it is no surprise to find excitability properties in ocean models. Weaver and Hughes \citep{Weaver94aa} discuss this effect in salt-conserving experiments with an idealised-geometry, ocean model. 
The ocean-atmosphere model of intermediate complexity CLIMBER (CLIMate BiosphERe mode) was shown to  exhibit excitability properties when boundary conditions are set to be typical of the latest glacial era \citep{Ganopolski02aa}.  The ocean circulation has then one stable state, with moderate Atlantic overturning, and a `quasi-stable state' with more intense overturning. The conceptual sketch of the excitation cycles shown by Ganopolski and Rahmstorf \cite{ganopolski01} on their Figure 1 can be interpreted in terms of slow-fast dynamics, in which the different states of the ocean circulation constitute the different branches of a slow manifold. The intense overturning state, which is the `plateau' phase of the Dansgaard-Oeschger event, may thus be viewed  as the repelling  branch of the slow manifold in the excitable regime (\figref{fig:sketch_oscillation}D). 
The  excitable Dansgaard-Oeschger hypothesis was used as a possible basis to explain how a weak forcing, exogeneous to the   system, could explain the observed 1500-yr periodicity of Dansgaard-Oeschger cycles  (on this periodicity: see ref. \citep{Schulz2002significance, Braun10aa} but see the other view in ref. \citep{Ditlevsen09ab}). Two such theories were developed on the basis of experiments with CLIMBER. One suggests that Dansgaard-Oeschger events are excited 
by stochastic fluctuations, modulated by a weak, hypothetical solar periodic forcing \citep{Ganopolski2002prl} (more on the effects of stochastic fluctuations in section \ref{sec:stoch}). The alternative theory suggests that the excitation is induced by the interference between two solar forcings with periods close to  $1470/7$ $(=210)$ and $1470/17$ $(\approx 87)$ years \citep{Braun05aa}, possibly combined with noise \cite{Braun08aa}. 

\subsubsection{Heinrich cycles as a relaxation oscillation.}
MacAyeal \citep{macayeal93heinrich} proposed an ice-binge/purge theory to explain Heinrich events. The theory rests on experiments with a 1-spatial direction model of ice flow dynamics. Suppose, as a starting point, that  ice volume grows in response to net accumulation of snow. The growth continues until the accumulated effect of geothermal heat flux causes basal sliding. A volume of ice is then released into the ocean (this is the `purge'), causing the release of icebergs characteristic of Heinrich events. Ice volume thus decreases, until ice accumulation wins over so that ice volume can grow again. The ice-binge/purge model is thus a relaxation oscillator combining a slow integrating process (ice mass accumulation) with a fast lateral discharge process.
\subsubsection{Coupling between Heinrich and Dansgaard-Oeschger events.}
To what extent Heinrich events may interfere with Dansgaard-Oeschger dynamics?
Paillard \citep{Paillard94aa,  Paillard95aa} investigated this question by coupling the MacAyeal ice model---but reduced to ordinary differential equations by Galerkin trun\-cation---with  a 3-box ocean model. The coupling simply assumes that ice released into the ocean causes a net freshening of the surface of the North Atlantic that alters the deep-ocean circulation. Paillard realised that this coupling could lead to fairly non-intuitive and complex effects, such as the succession of Dansgaard-Oeschger events of decreasing amplitude between Heinrich events. This succession is known in the litterature on palaeoclimate records as \emph{Bond cycles} \citep{bond92heinrich}. 
Paillard also found that the oscillations are aperiodic in this model under certain parameter configurations. 

The issue is further explored in \citep{Schulz2002oscillators},  based on the Winton-Sarachik ocean model, and  in \citep{Timmermann03ab}, based on the slightly more sophisticated Gildor-Tziperman ocean model \citep{Gildor-Tziperman-2000:sea}. The objective was to study the response of deep-decoupling ocean oscillations to prescribed Heinrich cycles. 
Schulz et al. \citep{Schulz2002oscillators} noted that deep-decoupling oscillations could be synchronised on the Heinrich cycles. Timmermann et al. \citep{Timmermann03ab} then proposed, on the basis of numerical experiments with a fairly idealised model, that  Heinrich events excite Dansgaard-Oeschger cycles because the variation in ice volume caused by a Heinrich event modifies slowly the amount of net freshwater released in the ocean. In turn, they suggested,  Dansgaard-Oeschger may have a control on ice volume growth. This yields a two-way coupling between Dansgaard-Oeschger and Heinrich events.

Experiments with more comprehensive models of the ocean-ice-sheet-atmosphere system \citep{Schmittner02aa, Ganopolski10aa} generally support the idea that the different water and heat fluxes involved in the different phases of ice build-up and iceberg release are quantitatively sufficient to support a coupling between ice sheets and ocean circulation during the latest glacial era. However, it was also noted that `` three-dimensional thermomechanical ice-sheet models are unable to satisfactorily reproduce the binge-purge mechanism without an \textit{ad hoc} basal parameterisation.'' \cite{Alvarez-Solas10aa}.
 
To address this difficulty a theory in which Dansgaard-Oeschger events trigger Heinrich events was recently proposed  \cite{Alvarez-Solas10aa}. The ice shelve plays a key role, in blocking the ice stream flow from the ice sheet to the oceans. Heinrich events occur when this ice shelve is broken, for example under the influence of ocean sub-surface warming associated with a Dansgaard-Oeschger event. The resulting model is a system displaying a slow ice-build-up -- Heinrich release cycle \textit{excited} by fluctuations in ocean sub-surface temperature. 
\subsubsection{Holocene oscillations and relationship with Dansgaard-Oeschger events.}
The much smaller ocean oscillations that characterised the Holocene period may also be a relaxation phenomenon. 
Schulz et al. \citep{Schulz07aa} observe oscillations in the atmosphere-ocean model of intermediate complexity ECBILT-CLIO. These oscillations are related to the convective activity in the Labrador Seas.  

Schulz et al. \citep{Schulz2004Glacial-intergl} considered the existence of such an oscillator in an earlier reference and speculated on the possible interactions between the centennial oscillations, millennial oscillations, and Heinrich cycles. They considered a model in which each of these three kinds of oscillations is modelled as a Morris-Lecar relaxation oscillator. Their working hypothesis is that glacial conditions induce a coupling between these oscillators. They then observed that a very stable 1500-yr oscillation appears, which they interpreted as a model equivalent of Dansgaard-Oeschger events. 

\section{Stochastic effects \label{sec:stoch}}
The myriad of chaotic motions that characterise the dynamics of the ocean and the atmosphere may be taken into account in the form of parameterisations involving stochastic time-processes. The method was introduced in climatology in the 1970s \citep{hasselmann76} and the theoretical justifications, which allow one to model chaotic motions as a (linear) stochastic process, are reviewed in ref. \cite{Penland03aa, Penland07aa}.  In a statistical inferential framework, the stochastic parameterisations may also be viewed as a way to account for the distance necessarily existing between the concepts and dynamics featured by the model, and the complex system being observed. 


The effects of stochastic processes on relaxation oscillators and excitable systems are generally well documented in the literature because this is a topic of general interest \citep{Lindner04aa}.  Here we review some of them in the specific context of palaeoclimate dynamics. 

\subsection{Stochastic effects on ice age dynamics}
\subsubsection{Phase dispersion.}
\begin{figure}[t]
\begin{center}
\includegraphics[width=0.9\textwidth]{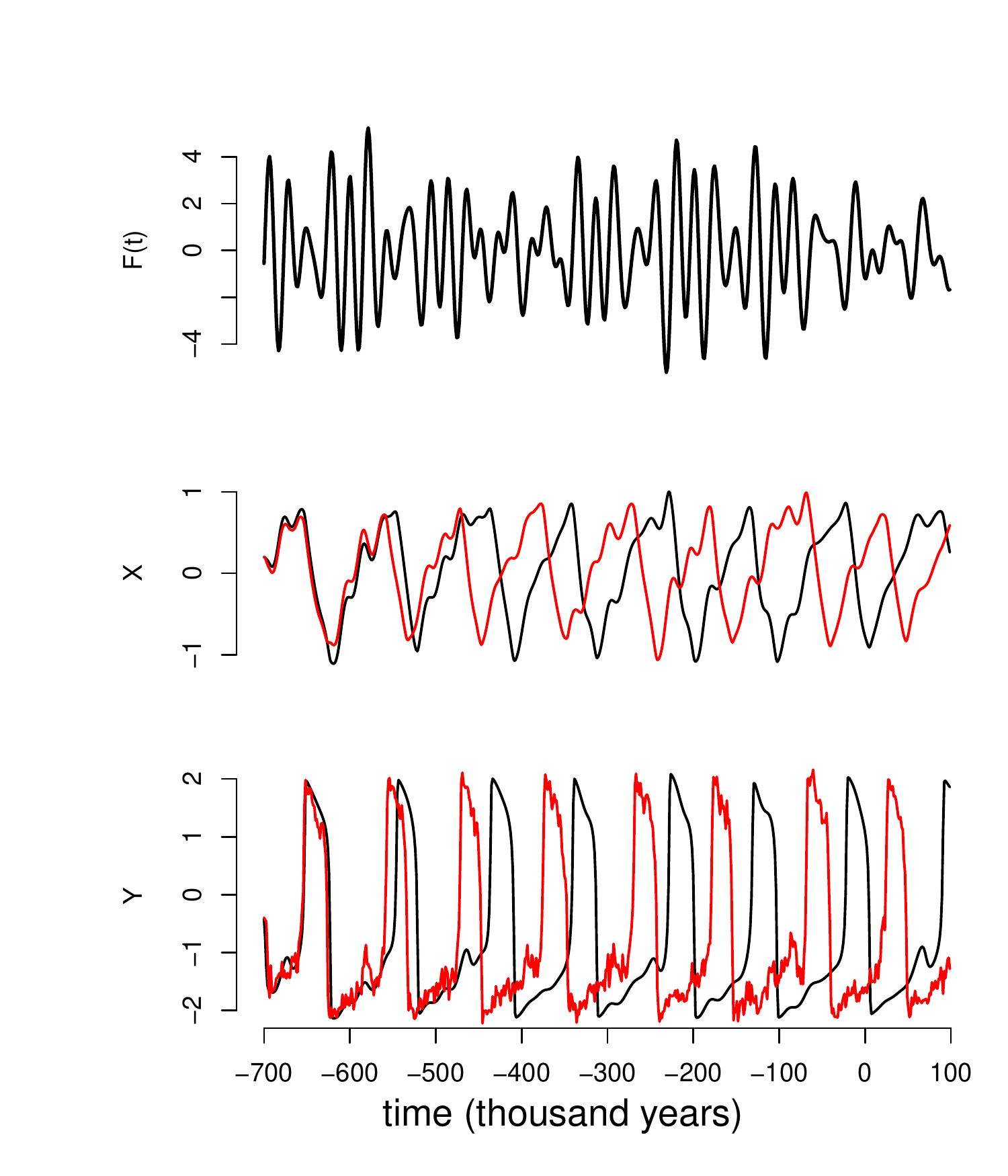}
\end{center}
\caption{The phenomenon of trajectory shift illustrated with the model (VDP) with astronomical forcing, with parameters $\tau=36\,ka$, $\beta=0.75$, $\gamma=0.4$, $\alpha=30$.  Black: the deterministic system. The generated history reproduces reasonably well the fluctuations of ice volume of the last 800,000 years. Red: same system but with an additive Wiener process added to the fast variable $y$, with variance $b = 0.2 (\mathrm{ka})^{-1/2}$. One possible realisation of the stochastic system is shown. Observe the solution shift around 450,000 years ago. Depending on the realisation the shift may occur at different places. }
\label{fig:localinstability}
\end{figure}

One of the basic effects of noise on oscillators is the phenomenon of \textit{phase dispersion}: A weak stochastic forcing on an oscillator causes a fading out of the memory of the exact initial conditions, even though the gross structure of the oscillation visualised in the phase space is conserved. The phenomenon is well known and it is an immediate consequence of the neutral stability of the phase of a  free oscillator with respect to fluctuations. It was early suggested that this phenomenon of phase dispersion may concern ice ages \cite{Nicolis87aa}, but it is more commonly believed that ice ages are phase-locked on the astronomical forcing. This phase-locking should act against dispersion and permit a very long predictability horizon of ice ages. 
Though, a phenomenon of phase dispersion may happen in oscillators that are locked on a periodic forcing. A stochastic fluctuation may momentarily cause a burst of desynchronisation, called phase slip, during which the system is unhooked from its corresponding deterministic trajectory and attracted to another trajectory, which leads or lags the original one by one forcing period (ref. \citep{Pikovski01aa}, sect. 3.1.3).
The difference between phase diffusion in a free system and in a periodic-forcing-driven oscillator is that the diffusion effect has, in the latter, a quantum nature.  More formally, it is said that the stochastic forcing disperses the system states around the different attractors that are compatible with the forcing. 
In a work in preparation we suggest that the astronomically-forced climate system may satisfy the conditions for a similar phenomenon of phase dispersion to occur (B. De Saedeleer, M. Crucifix and S. Wieczorek, unpublished data, 2011).  Given that the astronomical forcing is aperiodic the description of the phenomenon requires a suitable theoretical framework, which relies on the notion of a `local \emph{pullback} attractor'. 
The equivalent of a phase slip is, in the aperiodic forcing context, a stochastic shift from one of the deterministic pullback attractors to another one. The phenomenon is illustrated based on experiments with the VDP model on \figref{fig:localinstability}. 

\subsubsection{Reduction of period.}
\begin{figure}[t]
\begin{center}
\includegraphics[width=0.8\textwidth]{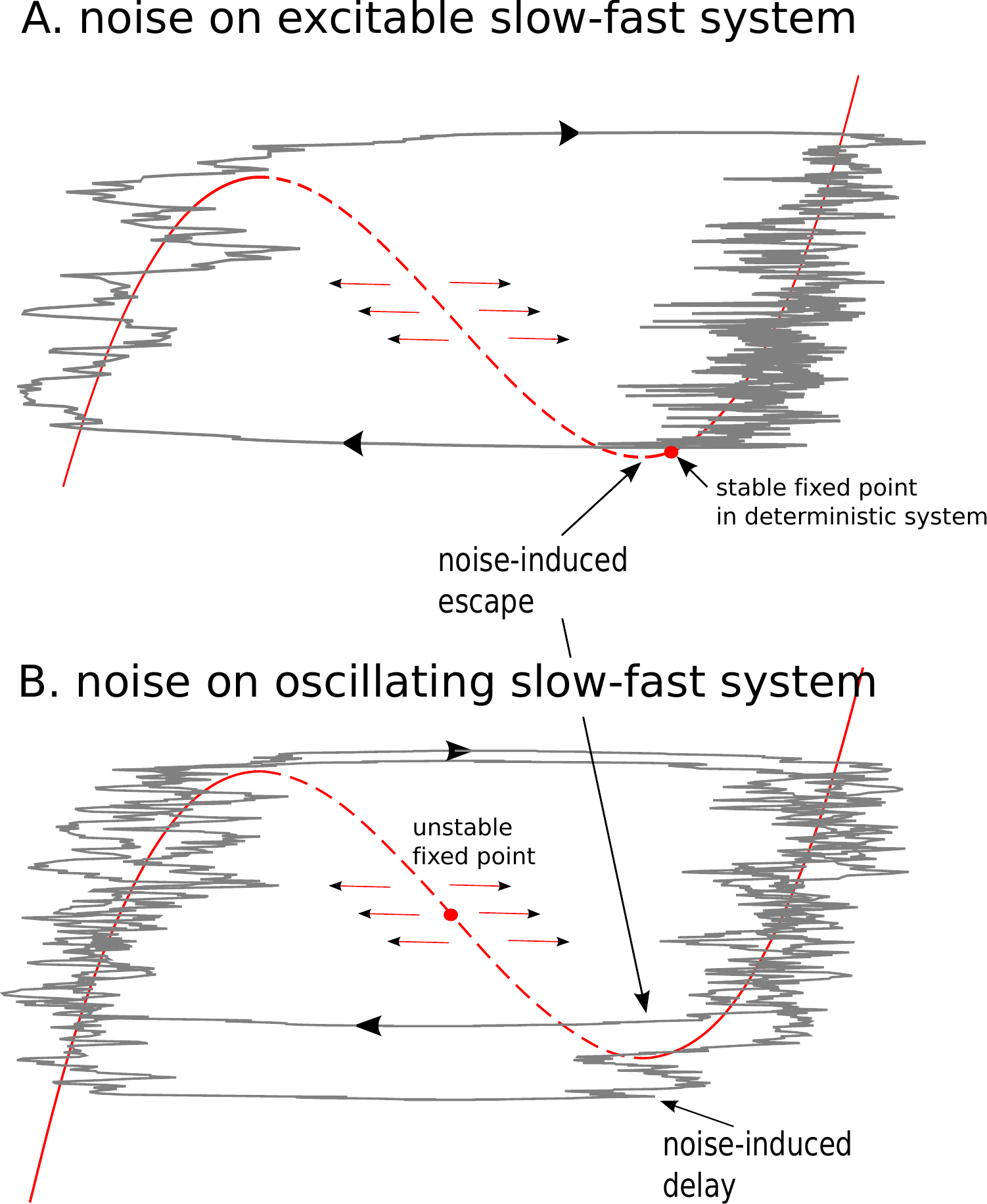}
\end{center}
\caption{
Two possible effects of additive noisy fluctuations in a slow-fast system. 
 \emph{(A.)} If the slow-fast system is in the excitable regime, noisy fluctuations can cause excitations. Excitations will be sporadic if the amplitude of the oscillations is weak (as shown here), or regular, as in a limit cycle, in the coherent resonance regime.  \emph{(B.)} If the slow-fast system is oscillating, additive fluctuations may cause early or delayed escapes from the slow branches, compared to the deterministic system. The net effet is a shortening of the cycle duration. 
 }
\label{fig:noisy_relaxation}
\end{figure}

Additive fluctuations generally reduce the period of relaxation oscillators. 
In an oscillator presenting a homoclinic orbit such as the Duffing oscillator, additive fluctuations reduce the time spent near the unstable focus \cite{Stone90aa}. This implies that even at the corresponding bifurcation point in the deterministic system, the return time of oscillations in the stochastically-perturbed system remains finite.  In a slow-fast oscillator such as the van der Pol oscillator, additive fluctuations generally result in early  escapes of the branch of the slow-manifold on which the system lies \figrefp{fig:noisy_relaxation}. The period of the oscillator is thus affected by a correction that increases approximately linearly with the noise variance in the slow-fast van der Pol oscillator \cite{Grasman89aa} \footnote{This result is established assuming fluctuations added to the slow variable. A much more general theory, suitable for different slow manifolds and additive fluctuations to slow and fast variables is now available \cite{Berglund02aa}.}. 
This property was used at least once in Pleistocene theory, in the silicate weathering hypothesis advanced by Toggweiler \cite{Toggweiler08ab}. Additive fluctuations reduce the limit-cycle period from 800,000 years to about 100,000 years. The reasons for the period reduction being so dramatic are left for another article. 

\subsection{Stochastic effects on Dansgaard-Oeschger dynamics}
\subsubsection{Stochastic excitation and resonance.}
Noise may naturally act as an excitation agent in an excitable system  \figrefp{fig:noisy_relaxation}. The topic is extensively reviewed in ref. \citep{Lindner04aa}. Excitation loops  are sporadic if the noise amplitude is weak, in which case the \emph{recurrence time} of the events is set by the noise amplitude, while the \emph{amplitude of the events} is set by the structure of the deterministic vector field. The frequency of excitation loops increases with the noise amplitude, until the system behaviour is qualitatively similar to a limit cycle regime. This is the coherent resonance regime. For yet higher noise amplitudes, the limit cycle structure is destroyed. 

The concept of stochastic excitation has been considered several times in Dans\-gaard-Oeschger theories. 
The idea is introduced based on experiments with an ocean general circulation model with idealised geometry and forcing \citep{Weaver94aa}. The effect of stochastic fluctuations is not only to excite oscillations in a system normally at rest, but also to reduce the period of these oscillations when the system is in oscillatory regime. 

A phenomenon of non-autonomous stochastic resonance may occur if the noise is superimposed to a weak external drive. For this to happen the autonomous system needs to be stable but excitable. The external forcing must be too weak to cause excitation by itself.  The role of the noise is to provide the additional power to induce excitation. The timing of the excitation is then related to the phase of the external forcing. The mechanism was proposed several times \cite{Timmermann03ab,Ganopolski2002prl}  to explain the 1500-yr recurrence time of Dansgaard-Oeschger events. 
The idea remains questioned, either on the ground that the 1500-yr recurrence time observed in palaeoclimate records is coincidental \citep{Ditlevsen09ab} or on the ground that the 1500-yr external forcing is unidentified \citep{Braun05aa, Colin-de-Verdiere07aa}. 
A more subtle case of stochastic resosonance involves the combination of noise with two solar cycles of 210 and 87 years, which yields the concept of `ghost resonance' \citep{Braun07aa} for which some support, albeit not conclusive, is found in the observations \cite{Braun10aa}.
\subsubsection{Decreased sensitivity to noise in resonant oscillators.}
Coupled oscillators may exhibit, collectively, a resonance period that is more robust to external fluctuations than the uncoupled oscillators. Schulz et al. \citep{Schulz2004Glacial-intergl} used this property to explain the stability of the Dansgaard-Oeschger recurrence period of 1500 years in presence of random fluctuations, without having to invoke an external forcing.
\subsubsection{Pseudo-oscillations in two-well systems.}
Finally, a behaviour reminiscent of oscillations may occur in a system that is neither oscillating nor excitable, but which presents several stable states. Noise then simply induces jumps between these different states. The simplest mathematical model is the Langevin equation and this is on this basis that Schulz et al. \citep{Schulz07aa} interpret the Holocene oscillations observed in the ECBILT-CLIO climate model. 

\section{Concluding discussion~: Can dynamical systems be used for inference? \label{sect:concl}}
The review  has shown that relaxation oscillations are a popular and powerful model to explain  oscillations observed in the Pleistocene record.  The concept of \textit{relaxation} implies some form of slow-fast separation, in the sense that at least one component of the system spends most of its time  in `quasi-equilibrium' states (this may be a `slow manifold branch' or a region influenced by a saddle-point, depending on the system structure), with acceleration phases. 

Some of these models were constructed following a fairly careful procedure of truncation of a system of partial differential equations, which describes some of the fluid dynamics of the climate system. Others were proposed on a more conceptual basis, the idea being precisely to test a hypothesis based on palaeoclimate observations. The latter approach is sometimes criticised, on the ground that box-models, for example, cannot reasonably be taken as an adequate representation of the complex dynamics of the oceans \cite{Wunsch10aa}.

This leads us to the last question of this review: can dynamical systems be used for inference on palaeoclimates?  Inference implies that something is being learned by confronting a model to observations. This inference process may take the form of a calibration procedure (update our knowledge on parameters on the basis of observations) or a model selection procedure (which model, among different alternatives, explains the observations best). 

The position taken here is that there is not such a thing as an `attractor' of the climate system that is to be  `discovered'. The hope is that some of its modes of behaviour are sufficiently de-coupled from the rest of the variability to justify the fact simple dynamical systems may capture the fundamental dynamical properties of these modes, and we want to learn about these modes from palaeoclimate observations. 

The programme is challenging. 
Indeed, it was underlined that different physical assumptions may lead to dynamical systems with dynamical properties that are similar enough to produce a convincing visual fit on palaeoclimate data \citep{tziperman06pacing}. The message is largely echoed in the present review. The modeller's challenge is therefore to operate a model selection on more stringent criteria than just fitting some standard time series. For example, palaeoclimate observations may yield constraints on the bifurcation structure of the system. The Middle-Pleistocene Transition is an attractive test case in this respect. 

In a statistical inference process, the observations should be a \textit{plausible} outcome or realisation of the model.  This makes sense only if the model has a stochastic component, which describes its uncertainties,   limitations, and the noise that emerges from the chaotic motions of the atmosphere and oceans. 

Stochastic dynamical systems begin to be used for inference on palaeoclimate time series. In a  method called `potential analysis', the climate system is modelled as a Langevin equation, that is, the combination of a down-gradient drift with a Wiener additive process, and inference is made on the number of wells of the potential function \citep{Livina10aa}. The method was applied on  Pleistocene climate records, yielding the conclusion that the number of wells increased from 2 to 3 over the course of the Pleistocene \citep{Livina11aa}. 

However, our position so far has been to favour a Bayesian methodology, because it allows one to encode physical constraints in the form of prior distributions on model parameters. The Bayesian formalism is also naturally designed for  model calibration, selection, and probabilistic predictions. 

The fact is that Bayesian methods for selection and calibration of dynamical systems on noisy observations are only emerging. In a recent attempt we considered a particle filter for parameter and state estimation \cite{Crucifix09aa}. To be honest, there is ample room for progress. Whether the process of inference with simple dynamical systems on palaeoclimate data will lead new insight in this context still needs to be demonstrated. 

\section*{Acknowledgements}
This review benefited from stimulating discussions during the Isaac Newton Programme  \textit{Mathematical and Statistical Approaches to Climate Modelling and Prediction} held in Cambridge 
during the summer-autumn 2010. Guillaume Lenoir,  Bernard De Saedeleer  and Jonathan Rougier provided helpful comments on a first version of the article. Thanks are also due to Didier Paillard, Olivier Arzel, Alain Colin de Verdi\`ere, Andr\'e Paul and Sebastian Wieczorek for e-mail correspondence during this review. The author is Research Associate with the Belgian National Fund of Scientific Research. This research is supported by the FP7-ERC starting grant ITOP ERG-StG-2009-239604. The editor (Jan Sieber) and two reviewers are acknowledged. 
\bibliography{/Users/crucifix/Documents/BibDesk.bib}
\end{document}